\documentclass[fleqn,10pt]{wlscirep}
\usepackage[utf8]{inputenc}
\usepackage[T1]{fontenc}
\title{Hard Optimization Problems have Soft Edges}

\author[1,2,3*]{Raffaele Marino}
\affil[*]{raffaele.marino@unifi.it}
\author[3]{Scott Kirkpatrick}

\affil[]{Dipartimento di Fisica e Astronomia, Università degli studi di Firenze,  Via Giovanni Sansone, 1, 50019 Sesto Fiorentino (FI), Italy}
\affil[2]{Dipartimento di Fisica, Sapienza Università di Roma, P.le Aldo Moro 5, Roma, 00185, Italy}
\affil[3]{School of Computer Science and Engineering, The Hebrew University of Jerusalem, Edmond Safra Campus, Givat Ram, Jerusalem 91904, Israel}

\begin{abstract}
Finding a Maximum Clique is a classic property test from graph theory; find any one of the largest complete subgraphs in an Erd{\"o}s-R{\'e}nyi $G(N,p)$ random graph. 
We use Maximum Clique to explore the structure of the problem as a function of $N$, the graph size, and $K$, the clique size sought.  It displays a complex phase boundary, a staircase of steps at each of which $2 \log_2N$ and $K_{\text{max}}$, the maximum size of a clique that can be found, increases by $1$.  
Each of its boundaries has a finite width, and these widths allow local algorithms to find cliques beyond the limits defined by the study of infinite systems.  We explore the performance of a number of extensions of traditional fast local algorithms, and find that 
much of the "hard" space remains accessible at finite $N$. 

The "hidden clique" problem embeds a clique somewhat larger than those which occur naturally in a $G(N,p)$ random graph.  Since such a clique is unique, 
we find that local searches which stop early, once evidence for the hidden clique is found, may outperform the best message passing or spectral algorithms.

\end{abstract}

\begin{document}

\flushbottom
\maketitle
%
%
\thispagestyle{empty}

\section*{Introduction}

Phase transitions and phase diagrams describing the behavior of combinatorial problems on random ensembles are no longer surprising. 
Large scale data structures arise in practical examples, such as the analysis of large amounts of social data, extending to the activities of a few billion people. Effective tools for managing them have commercial value.  The model system and the forms of interactions that couple its elements in data science are known. While exact methods can solve only very small examples, approximate simulation of medium scale problems must reach very large scale. Methods such as 
finite-size scaling analysis expose regularities \cite{kirkpatrick1985statistical}.  Classic examples include the Satisfiability problem in its many variants \cite{selman1993local,kirkpatrick1994critical}. 

In this paper, we consider finding maximum cliques in random graphs, specifically Erd{\"o}s-R{\'e}nyi \cite{erdos1959random,bollobas1998book}  graphs of the $G(N,p)$ class, with $N$ nodes (or sites) and each edge (or bond) present with probability $p$.   We further specialize to the case $p = 1/2$.
\begin{figure}
\centering
\includegraphics[width=1\columnwidth, keepaspectratio=true, angle=0]{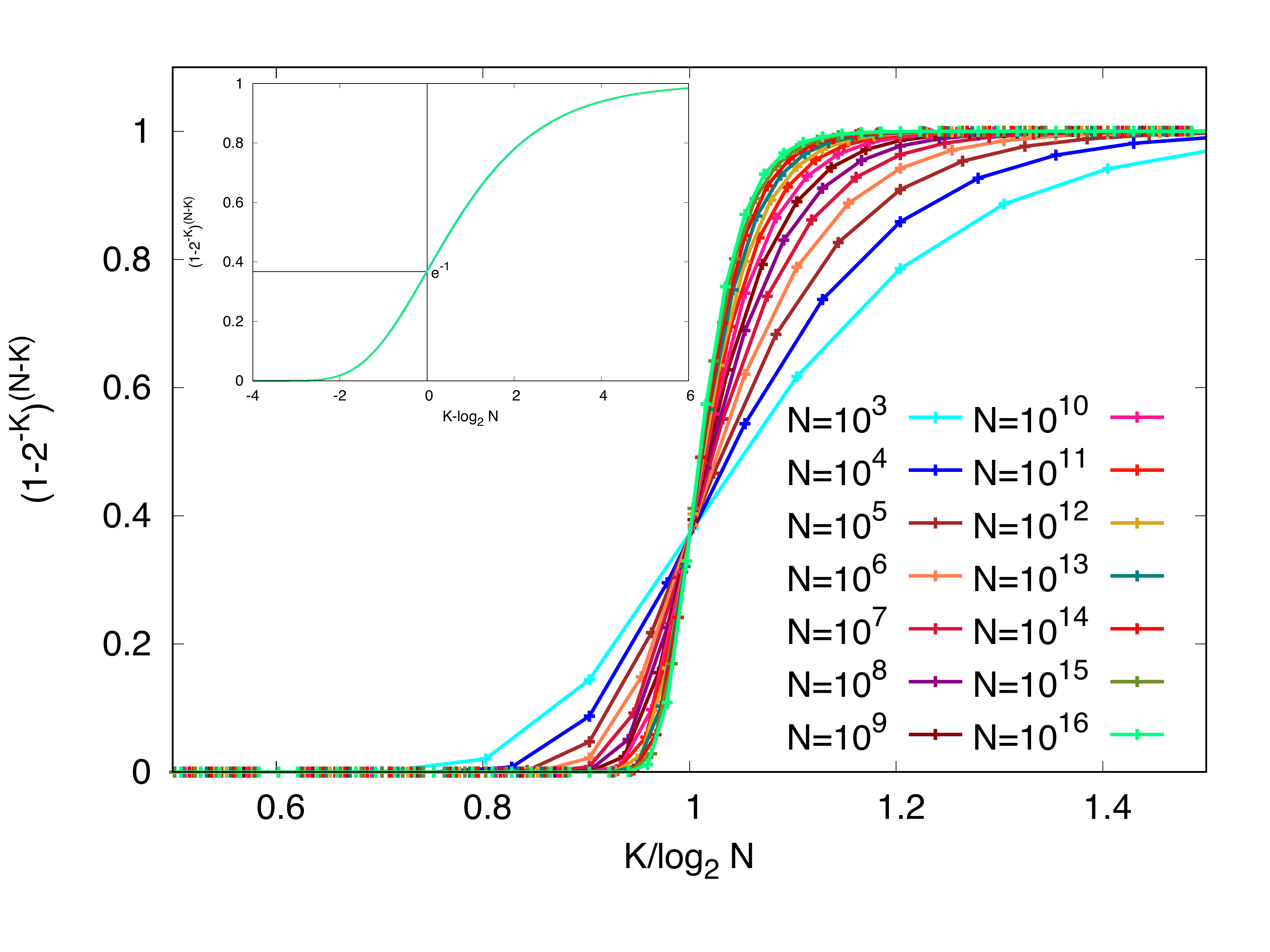}
\caption{The picture shows the probability that we can find no other site to grow the clique to size $K+1$ as a function of $K/\log_2 N$. 
The inset shows the universal form that all of these curves take when rescaled.}
\label{figexpanding}
\end{figure}
Maximum Clique is an unusually difficult problem,
for which naive solution methods (Fig.\ref{figexpanding}) can construct cliques of size $K=\log_2 N$, yet probabilistic arguments show that solutions asymptotically of size $2\log_2 N$ must exist.  No polynomial algorithms that will construct true maximum cliques for arbitrarily large values of $N$ are known. The failure is general, not merely a problem for the rare worst case.  
We shall test several fast algorithms of increasing complexity to determine the range of $N$ at which each gives useful answers.

Greedy methods are fast, and a good starting point for our discussion. 
Start with a site anywhere in the graph, and discard roughly half of the sites that are not neighbors. 
Pick a neighbor from the \textit{frontier} that remains.  Then discard half of the remaining sites that are not a neighbor of the new site.  
Continue in this way until the \textit{frontier} vanishes -- no candidates to extend the clique remain.  
Since we have halved the size of the frontier at each step, it is unlikely that this process can proceed beyond $\log_2 N$ steps \cite{karp1976probabilistic}.

Let's look at this more precisely, as a function of $N$. 
The stopping probability, that we can find no other site to grow from size $K$ to $K+1$ is $(1-2^{-K})^{(N-K)}$, 
as shown in Fig.\ref{figexpanding}, where use of a common scale $K/\log_2 N$, brings the various curves all together at a probability of $e^{-1}$ when $K$ is equal to $\log_2 N$.  All cliques are extendable when $K<< \log_2 N$, and none are when $K>>\log_2 N$. The slopes of these curves are each proportional to $\log_2 N$.  The simple expedient of plotting the curve for each value of $N$ against $K - \log_2 N$ collapses all of them to a universal limiting form, which is shown in the inset to Fig. \ref{figexpanding}. This is finite-size scaling just as seen in phase transitions\cite{kirkpatrick1985statistical}. It also shows that cliques selected at random from the large number which we know must exist start to be non-extendable at a size two sites below $\log_2 N$ boundary and are almost never extendable four sites above, with a functional form that is almost independent of $N$.  This threshold occurs for each $N$ at $K = \log_2 N$, and has a width in $K$ which is independent of $N$.


Matula first called attention to several interesting aspects of the Maximum Clique problem on $G(N,p)$.  From the expected number of cliques $n(K,N)$, of size $K$ at $p = 1/2$ \cite{matula1970complete}:

\begin{equation}
\label{eq::averageclique}
n(K,N) = {N \choose K} 2^{-{K \choose 2}} ,
\end{equation} 

using Stirling's approximation, one can see that this is large at $K = \log_2 N$ but becomes vanishingly small for $K > 2\log_2 N$, providing an upper bound to $ K$.  Matula identified $K_{\text{max}}$ as the largest integer such that 
\begin{equation}
\label{eq::KMAX}
n(K_{\text{max}},N ) \geq 1 .  
\end{equation}
and \cite{matula1970complete,matula1972employee}  expanded the finite $N$ corrections to the continuous function $R(N)$ solving $n(R,N) = 1$:

\begin{equation}
R(N)=2\log_2 N-2\log_2 \log_2 N+2 \log_2 (e) .
\label{MatBoleqlong}
\end{equation}

\begin{figure}
\centering
\includegraphics[width=1\columnwidth, keepaspectratio=true, angle=0]{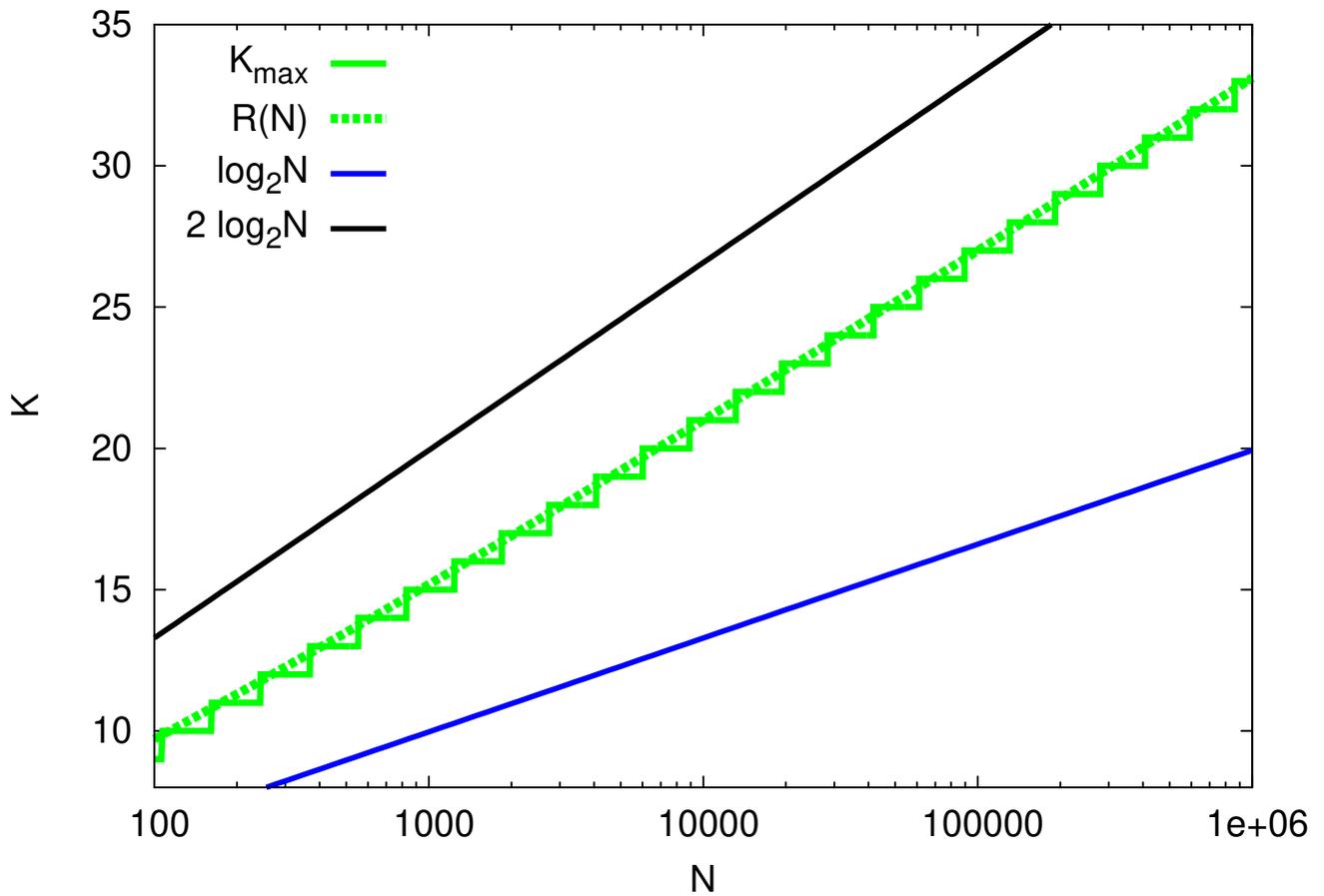}
\caption{Maximum size of a clique in $G(N,0.5)$ as function of the order of the graph. In green are presented $K_{\text{max}}$, result of  solving (\ref{eq::KMAX}) and $R(N)$ given by  (\ref{MatBoleqlong}), respectively. In black is plotted the value of $2 \log_2 N$ and in blue, $\log _2 N$. }
\label{Matstaircase}
\end{figure}

\begin{figure}
\centering
\includegraphics[width=1\columnwidth, keepaspectratio=true, angle=0]{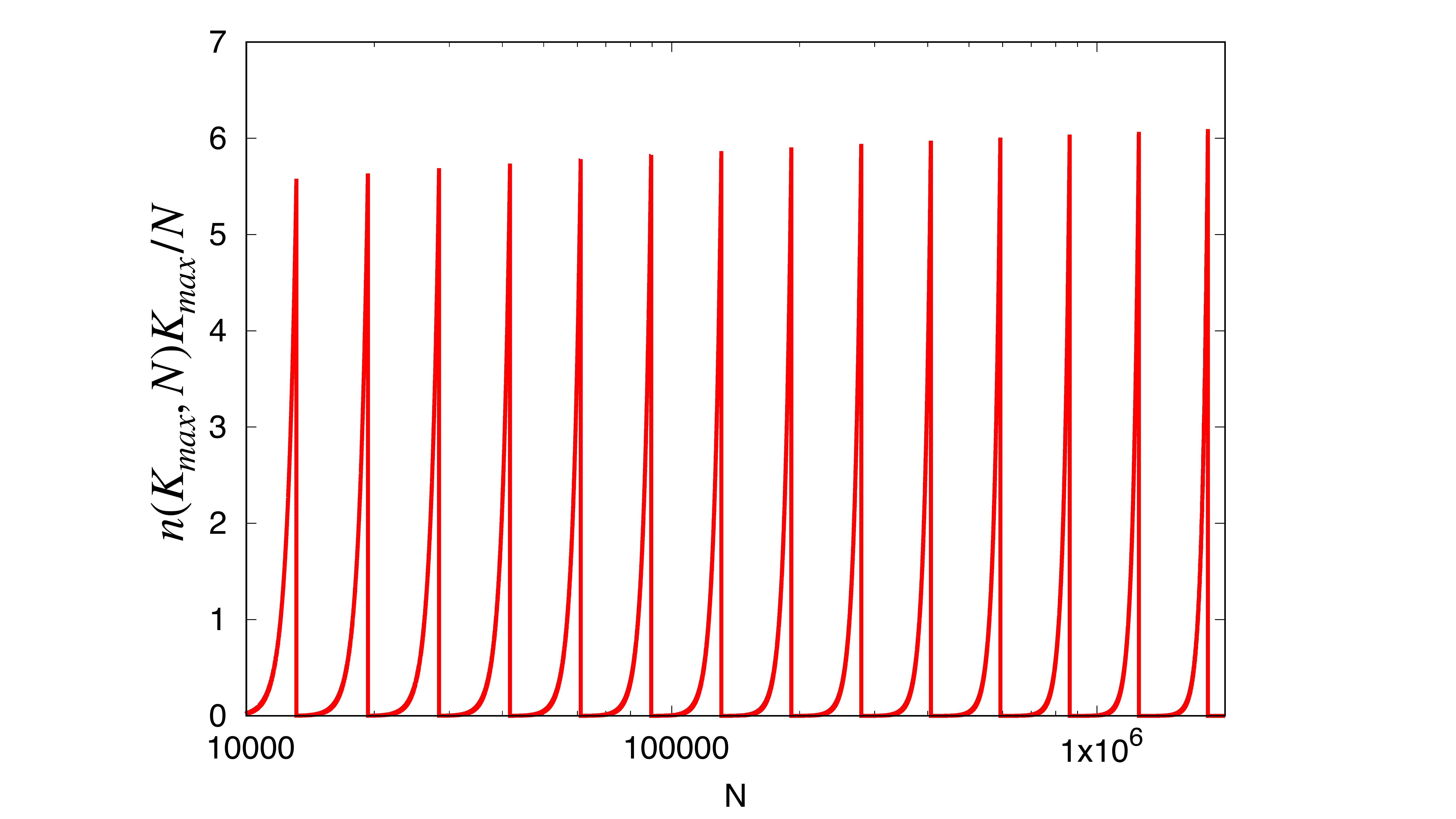}
\caption{Expected number of maximum cliques of size $K_{\text{max}}$ per site in the graph, as a function of N}
\label{fig-xx}
\end{figure}

This formula is also discussed in Bollob{\'a}s and Erd{\"o}s \cite{bollobas1976cliques} and by Grimmett and McDiarmid \cite{grimmett1975colouring}. 
We will focus on $K_{\text{max}}$, the predicted actual maximum clique size.  In effect, its value follows a staircase with the prediction (\ref{MatBoleqlong}) passing through the risers between steps, as shown in Fig. \ref{Matstaircase}.  Across each step $n(K_{\text{max},N})$ grows from $1$ to $\mathcal{O}(N)$.  The number of maximum cliques per site in the graph that results is shown in Fig. \ref{fig-xx}. We see that at the left edge of each step, cliques of size $K_{\text{max}}$ are very rare, but at the right edge of that step, each site is possibly contained in multiple maximal cliques, while cliques of size $K_{\text{max}} + 1$ have an expectation which has reached $1$.  Thus across each step, the probability of finding a clique of larger $K$ decreases by a factor of  $\mathcal{O}(N)$ for each increase by 1 in $K$.
We will focus on $K_{\text{max}}$, the predicted actual maximum clique size.  In effect, its value follows a staircase with the prediction (\ref{MatBoleqlong}) passing through the risers between steps, as shown in Fig. \ref{Matstaircase}.  Across each step $n(K_{\text{max}})$ grows from $1$ to $\mathcal{O}(N)$.  The number of maximum cliques per site in the graph that results is shown in Fig. \ref{fig-xx}. We see that at the left edge of each step, cliques of size $K_{\text{max}}$ are very rare, but at the right edge of that step, each site is possibly contained in multiple maximal cliques, while cliques of size $K_{\text{max}} + 1$ have an expectation which has reached $1$.  Thus across each step, the probability of finding a clique of larger $K$ decreases by a factor of  $\mathcal{O}(N)$ for each increase by 1 in $K$.

\begin{figure}
\centering
\includegraphics[width=1\columnwidth, keepaspectratio=true, angle=0]{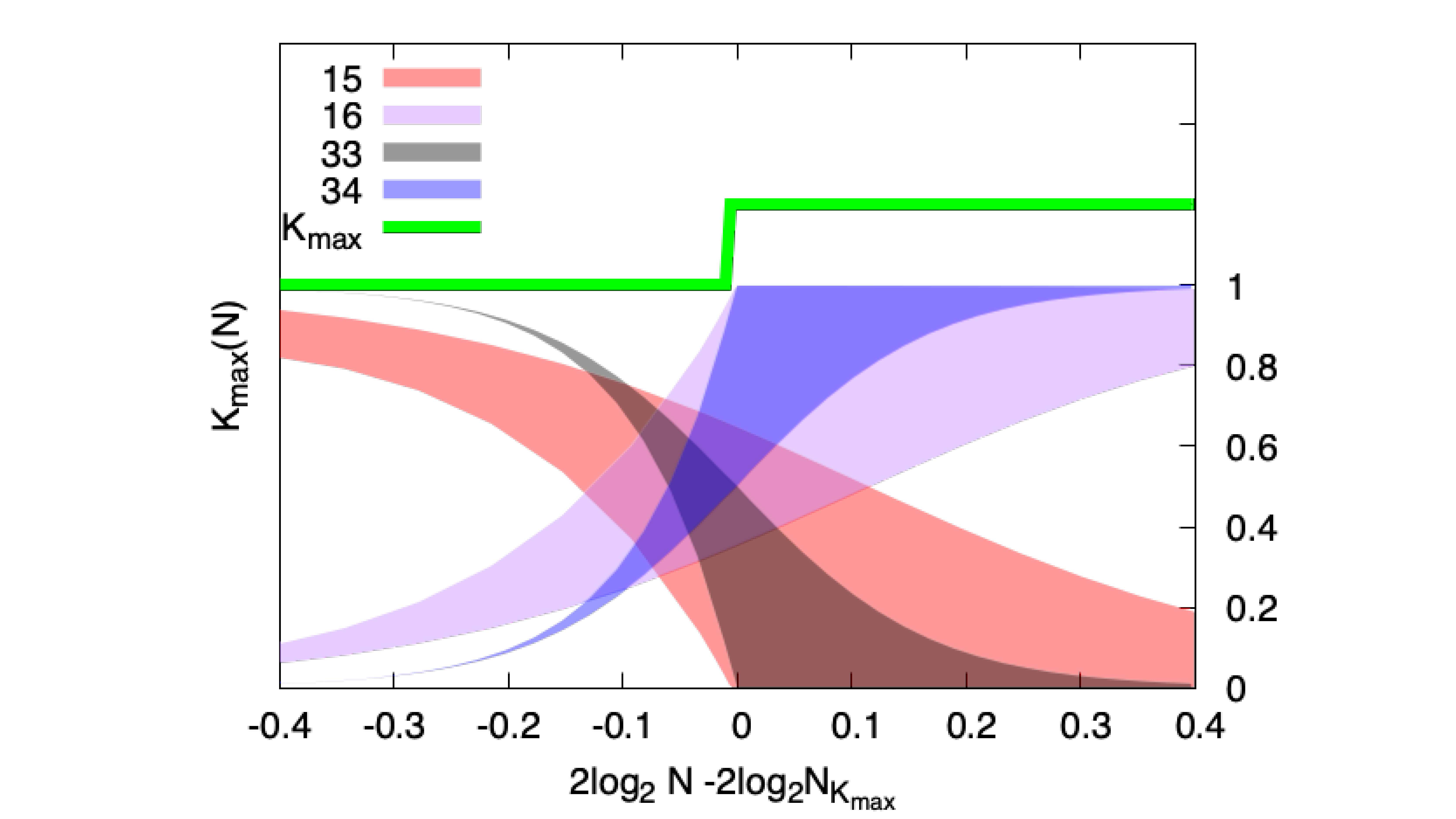}
\caption{The picture shows regions defined by upper and lower bounds (obtained by equation (\ref{eq::2})), on the fractions of graphs of size $N$ with maximum cliques of size $K_{\text{max}}$ and $K_{\text{max}} + 1$. The steps described occur at $N = 1239$ (the lighter colors bound the fraction of cliques of size 15 and 16) and $N = 1254516$ (the darker colors bound the fractions of cliques with sizes 33 and 34). The figure depicts approximately half a step width to either side of the transition for each size.} 
\label{bigMatprob}
\end{figure}

Matula drew attention to a \textit{concentration} 
result for the clique problem.  As $N \to \infty$ the sizes of the largest cliques that will occur are concentrated on just two values of $K$, the integers immediately below and above $R(N)$.  From the second moment of the distribution of the numbers of cliques of size $K$ one can bound the fraction of graphs with no such cliques, and sharpen the result \cite{matula1976largest} by computing a weighted second moment.    In effect, Markov's inequality provides upper bounds, and Chebyscheff's inequality provides lower bounds on the existence of such cliques.  
The probability that the maximum clique size is $K_{\text{max}}$ was given by Matula  \cite{matula1970complete,matula1972employee,matula1976largest}.  The fraction of graphs $G(N,p)$ with  maximum  clique size $K_{\text{max}}$, is bounded as follows:
\begin{equation}
\label{eq::2}
\resizebox{0.43\textwidth}{!}{$\left(\sum_{j=\text{max}\{0, 2k-N\}}^k \frac{{N-k \choose k-j}{k \choose j}}{{N \choose k}} p^{-{j \choose 2}}  \right)^{-1}\leq \text{Prob}(K_{\text{max}}\geq k) \leq {N \choose k} p^{k \choose 2}.$}
\end{equation}
This leads to the following picture, evaluated for  
two values of $N$, one small and one quite large, in Fig. \ref{bigMatprob}, 
we see that at the step between two integer values of $K_{\text{max}}$, more than half of the graphs will have a few cliques of the new larger value from the upper step, and less than half will have only cliques with the smaller value from the lower step, but many of them.  Fig. \ref{bigMatprob} shows that the crossover at each step edge narrows with increasing $N$, but only very slowly.  The two cases sketched correspond to steps at roughly $N = 1.2\, 10^3$ and $N = 1.3 \, 10^6$. For the case at smaller $N$, the transition is spread over almost half the width of the step. As in Fig.
\ref{figexpanding}, the transition is not symmetric.  The appearance of cliques with the new value of $K_{\text {max}}$ is sharper and comes closer to the step than the disappearance of cases in which the cliques from the previous step still dominate. Because the natural scale of this problem is $\log_2 N$, we see that the width of the transition from one step to the next remains important on the largest scales encountered in actual data networks, such as $N \sim 10^9$, the population of the earth.


The traditional approach to surveying and challenging the developers of algorithms for solving hard problems is to assemble a portfolio of such problems, some with a known solution, and some as yet unsolved.  The DIMACS program at Rutgers carried out such a challenge in the mid $1990$'s \cite{johnson1996cliques}.  Roughly a dozen groups participated over a period of a year or more, and the sample graphs continue to be studied.  The largest graphs in the portfolio were random graphs of size $1000$ to $2000$, and the methods available gave results for these which fell at least one or two short of $R(N)$.  As a result, the actual values of $K_{\text{max}}$ for many test graphs are still unknown.   A better test for these algorithms on random graphs is to determine to what extent they can reproduce the predicted distribution of results that we see in Fig. \ref{bigMatprob}, both the steps in $K_{\text{max}}$ and the fraction of graphs with each of the dominant values of $K_{\text{max}}$ as it evolves with increasing $N$.  We will present extensions of the algorithms tested at DIMACS and show that they give good results on sets of problems larger than those in the DIMACS portfolio.

Several authors have proposed that the search for powerful, effective clique-finding algorithms could be expressed as a \textit{challenge}.
Mark Jerrum, in his $1992$ paper \textit{``Large Cliques Elude the Metropolis Process''}, \cite{jerrum1992large} sets out several of these. 
His paper shows that a restricted version of stochastic search is unlikely to reach a maximum clique, and also introduces the additional problem of finding an artificially \textit{hidden clique}, which we discuss in a later section.  A hidden clique or \textit{planted solution}, is just what it sounds like, a single subgraph of $K_{HC}$ sites, with $K_{HC} > K_{\text{max}}$, so that it can be distinguished, for which all the missing bonds among those sites have been restored. 
A series of papers \cite{alon1998finding,dekel2014finding} show that if $K_{HC}$ is of order $N^{\alpha}$ with $\alpha > 0.5$, a small improvement over our naive greedy algorithm ($SM^{0}$ introduced in the next section) will find such a hidden clique, simply by favoring sites in the frontier with the most neighbors in its search. 

Jerrum's first challenge is to find a hidden subgraph of size $\sim N^{0.5 - \epsilon}$ with probability greater than $ 1/2$, using an algorithm whose cost is polynomial in the number of bonds in the graph (i.e. $N^2$ is considered to be a linear cost).  We see below that the range of hidden clique sizes that can be hidden between $N^{1/2}$ and the naturally-occurring cliques of size 
$2 \log_2N$ is not large.  So practically oriented work has focused on finding hidden cliques of size $\alpha N^{1/2}$ for $\alpha < 1$.  Karp in his original paper, Jerrum, and several others have also turned the identification of any naturally occurring clique larger than $\log_2 N$ into such a challenge: find any clique of size exceeding $\log_2 N$ with probability exceeding $1/2$.  We saw in the discussion of Fig.\ref{figexpanding} that finding cliques, which exceed $\log_2 N$ by a small constant number of sites should be straightforward at any value of $N$. 
We shall see that both challenges can be met for large, finite, and thus interesting values of $N$, and will attempt to characterize for what range of $N$ they remain feasible.

Below, in section 1, we introduce a class of greedy algorithms, of  complexity polynomial in $N$, and test the accuracy with which they reflect the size and distribution of the maximum cliques present in $G(N,p=1/2)$ for useful sizes of $N$.  In section 2, we review the "hidden clique" variant of the Maximum Clique problem, and describe its critical difference from the problem treated in Section 1 (that the hidden clique is unique, while naturally occurring cliques are many). 
We show that our greedy algorithms can exploit this difference to perform as well as or better than the best proposed methods in the literature. Section 3 summarizes our conclusions and provides recommendations for future work.

\section{Greedy Algorithms}
\label{sec:maximumcliqueproblem}

In this section, we describe the performance of a family of increasingly powerful greedy algorithms for constructing a maximal clique on an undirected graph.  Those algorithms are polynomial in time and use some randomness, but they are myopic in generating optimal solutions. However, because they are relatively fast, significant research effort has been devoted to improving their performance while adding minimal complexity. We will show ways of combining several of these simple greedy algorithms, to obtain better solutions at somewhat lower cost by adding a very limited form of back-tracking.

We start by considering a simple family of greedy algorithms, designated by Brockington and Culberson \cite{brockington1996camouflaging}, as $SM^{i}$, $i=0,1,2..$. 
$SM^{0}$ improves over the naive approach we described at the outset \cite{kuvcera1991generalized}, by selecting at each stage the site with the largest number of neighbors to add to the growing clique. If there are many such sites to choose from, each connected to all of the sites in the part of the clique identified to that point, one is chosen at random, so multiple applications of $SM^{0}$ will provide a distribution of answers for a given graph $G(N,p)$. At each stage this choice of the site to add retains somewhat more than half of the remainder of the graph, $Z$, so the resulting clique will be larger than $\log_2 N$, for all $N$.  $SM^0$ can be implemented to run in $\mathcal{O}(N^2)$ time.


$SM^{i}$ for $i=1, 2,...$ are algorithms in which we start our greedy construction with each combination of $i$ vertices which form a complete subgraph, then extend them one site at a time using $SM^0$. In other words, $SM^{0}$ is run starting with each of ${N \choose i} p^{{i \choose 2}}$ complete subgraphs of order $i$. $SM^{1}$, starting with every site, can be implemented to run in $\mathcal{O}(N^3)$. The complexity of $SM^{2}$, which uses all connected pairs, is $\mathcal{O}(N^4)$. The computational complexity for the class of algorithms $SM^{i}$ is $\mathcal{O}(N^{i+2})$.

\begin{figure}
\centering
\includegraphics[width=1\columnwidth, keepaspectratio=true, angle=0]{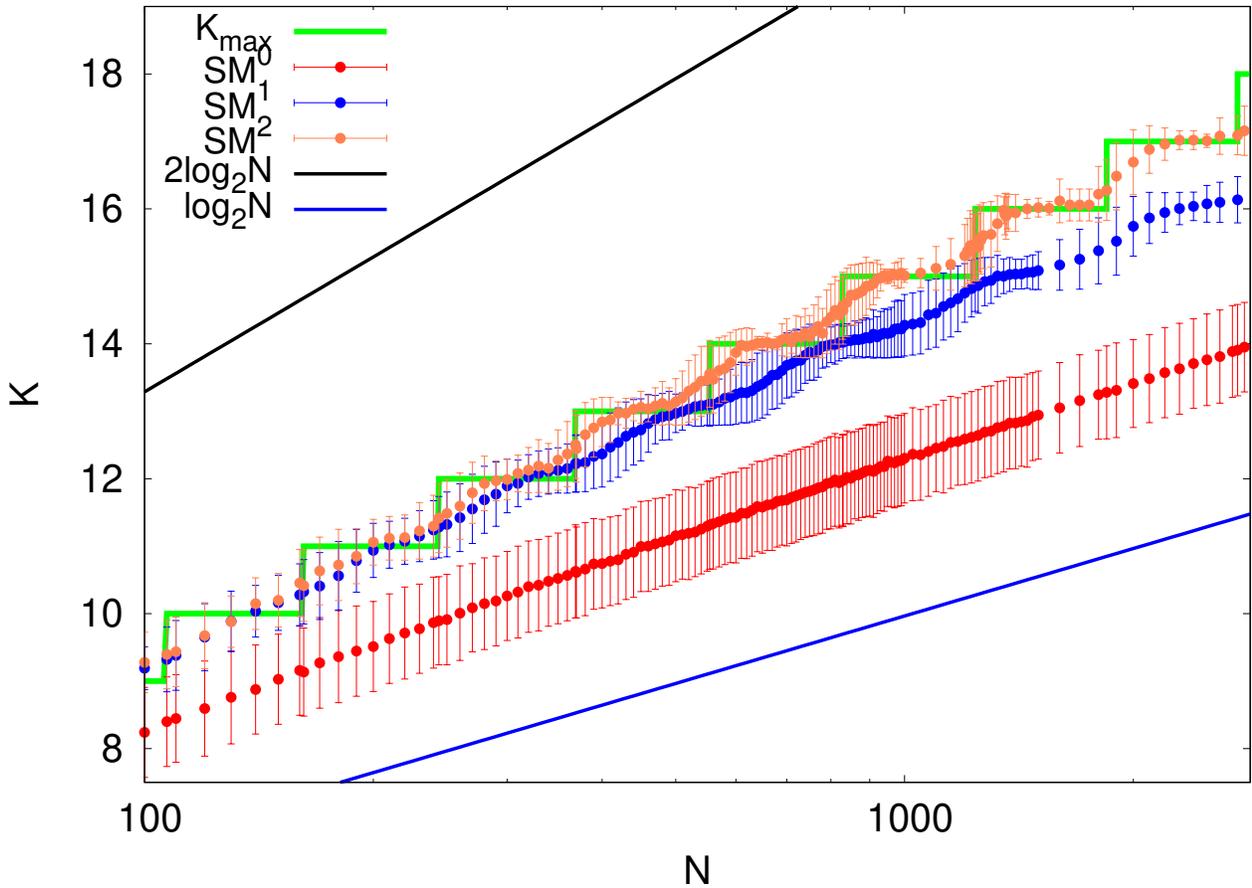}
\caption{Mean and standard deviations of sizes of the maximal cliques from Erd{\"o}s-R{\'e}nyi graphs $G(N, p=0.5)$, using the algorithms $SM^{i}$, with $i=0, 1, 2$. The maximum size of a clique as a function of the order of the graph is represented by the green staircase $K_{\text{max}}$. The black and the blue straight lines show $2\log_2 N$ and $\log_2 N$, respectively. Red data points describe the mean maximal clique size obtained by $SM^{0}$, averaging over $2000$ Erd{\"o}s-R{\'e}nyi graphs. Blue points describe the maximal clique size found with $SM^{1}$, while orange points are the results of $SM^{2}$. The average is obtained over $500$ and $100$ graphs for $SM^{1}$ and $SM^{2}$, respectively. }
\label{figSM0SM1SM2}
\end{figure}

In Fig. \ref{figSM0SM1SM2} we show the sizes of the maximal cliques on Erd{\"o}s-R{\'e}nyi graphs $G(N, p=0.5)$, found using the algorithms $SM^{i}$, with $i=0, 1, 2$. For comparison, we plot the green staircase, $K_{\text{max}}$.
This figure shows the improvements that result from the (considerable) extra computational cost of the latter two algorithms.  Both the blue points of   $SM^{1}$ and the orange points of $SM^{2}$ reflect the staircase of $K_{\text{max}}$. Even their error bars reflect the rapid increase of the number of the larger maximum cliques after each jump in the staircase.  The error bars on the red points of $SM^{0}$ do not show any staircase pattern. $SM^{0}$, even with error bars,  lies consistently above $\log_2 N$. Each red point is the average over $2000$ random Erd{\"o}s-R{\'e}nyi graphs, each blue point is the average over $500$ random Erd{\"o}s-R{\'e}nyi graphs, and each coral point is the average over $100$ random Erd{\"o}s-R{\'e}nyi graphs. We have used a uniform random number generator with an extremely long period (WELL1024) \cite{panneton2006improved}.  The $SM^{2}$ results track the staircase closely up to $N = 4000$, a larger size than seen in the DIMACS study, while the $SM^{1}$ results fall about $1$ site below the staircase at the end of this range.

\begin{figure}[t]
\centering
\includegraphics[width=1\columnwidth, keepaspectratio=true, angle=0]{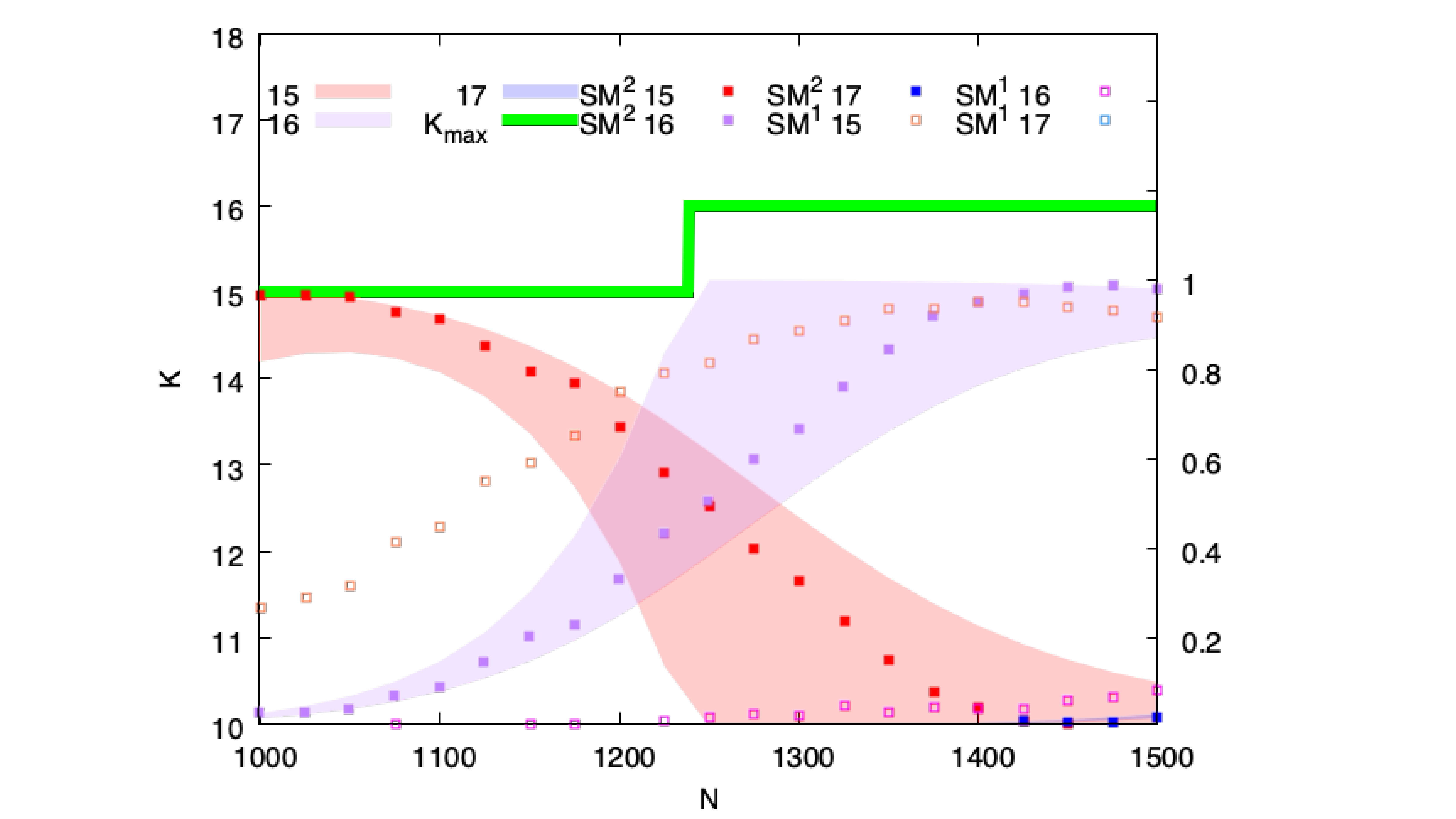}
\caption{ Across a single step in $K_{\text{max}}$, this figure compares the fraction of graphs predicted to have each value of $K_{\text{max}}$ from equation (\ref{eq::2}) with the experimental probability obtained using $SM^1$ and $SM^2$. The left $y$-axis shows the clique number, i.e. the largest clique size $K_{\text{max}}$ as a function of $N$. The right $y$-axis indicates the fraction of $K_{\text{max}}-1$, $K_{\text{max}}$, $K_{\text{max}}+1$, obtained from the experiments with $SM^1$ and $SM^2$. Each fraction has been computed on a population of $500$ random Erd{\"o}s-R{\'e}nyi graphs $G(N,p=0.5)$. The regions coloured in red, purple, and blue identify the theoretical probability obtained by equation (\ref{eq::2}). The filled square points describe the fraction of $K_{\text{max}}-1$, $K_{\text{max}}$, $K_{\text{max}}+1$ (red, purple, and blue respectively) obtained from experiments with $SM^2$, the empty squares the same for $SM^1$. 
}
\label{figSM1SM2}
\end{figure}

Fig. \ref{figSM1SM2} gives a more detailed comparison of the two algorithms.  Fig. \ref{figSM1SM2} compares the predicted fraction of random Erd{\"o}s-R{\'e}nyi graphs having a maximal clique size with the experimental results obtained with the two algorithms around the step from $K_{\text{max}} = 15$ to $16$. This corresponds to the region most often explored in the DIMACS studies.
The $SM^{2}$ algorithm remains within the bounds described by Matula.  The red, purple, and blue filled square points in the three predicted probability regions (red, purple, and blue, respectively) find acceptable fractions of $15$, $16$, and even $17$ sites cliques as $N$ is increased.  The $SM^{1}$ algorithm, shown by  orange, pink, and blue open squares, falls short in all three probability regions, finding too many $15$'s, too few $16$'s, and no $17$'s. The two algorithms both provided similar distributions of results up to $N = 300$ of $K_{\text{max}} = 12$.


These two algorithms are very expensive. We could only analyze rather small random graphs, comparable to the larger DIMACS examples. The results fall within the calculated distribution, which is a valuable check, since the actual value of $K_{\text{max}}$ is unknown for an individual graph.  Next, we consider less costly algorithms, which allow us to explore much larger graphs. These give results lying between $SM^{0}$ and $SM^{2}$ and still reflect the staircase character of the underlying problem.

We reverse the order of operations made by the class of algorithm $SM^{i}$, with $i=1, 2, ..$. Instead of running $SM^{0}$ for each pair, or triangle, or tetrahedron (etc.) in the original graph, we run $SM^{i}$, with $i=1, 2, ..$ fixed, but only on the sites found within one solution given by $SM^{0}$.  $SM^0$ will return a clique $C$ of size $|C|$. On this solution we run $SM^i$, i.e. we select all the possible ${|C|\choose i}$ complete subgraphs in the clique $C$, and, on each of them, we run $SM^0$. 
This simple algorithm, which we call $SM^0 \to SM^i$, will run in a time bounded by $\mathcal O(N^2 \ln N)$.
 




We have analyzed the results of the algorithm $SM^{0}\to SM^{i}$, with $i$ fixed to $4$,  compared to  $SM^0$ in the range of $N$ $[2800:50000]$. Thus we analyzed ${|C|\choose 4}$ graphs of order approximately $N/16$. The combined algorithm $SM^{0}\to SM^{4}$ always finds a maximal clique bigger than those given by $SM^0$ alone. Moreover, the combined algorithm reproduces the wiggling behaviour due to the discrete steps in $K_{\text{max}}$ in a time bounded by  $\mathcal{O}(N^2 \ln N)$, while $SM^{0}$, used alone, does not.

The improved results of the combined algorithm $SM^{0} \to SM^{i}$, with fixed ${i}$, suggests iterating the procedure. First we run $SM^{i}$, with fixed $i$, on the clique returned by $SM^{0}$. If the clique returned by the algorithm is bigger than the one that is used for running $SM^{i}$, then we use the new clique as a starting point where $SM^{i}$ will be run again. The algorithm stops when the size of the clique no longer increases. The complexity of the algorithm, therefore, is  $\mathcal{O}(tN^2 \ln N)$, where $t$ is the number of times we find a clique, which is bigger than the previous one. We call, thus, this new algorithm $SM^{0}\to \text{iter}[SM^{i}]$.

\begin{figure}
\centering
\includegraphics[width=1\columnwidth, keepaspectratio=true, angle=0]{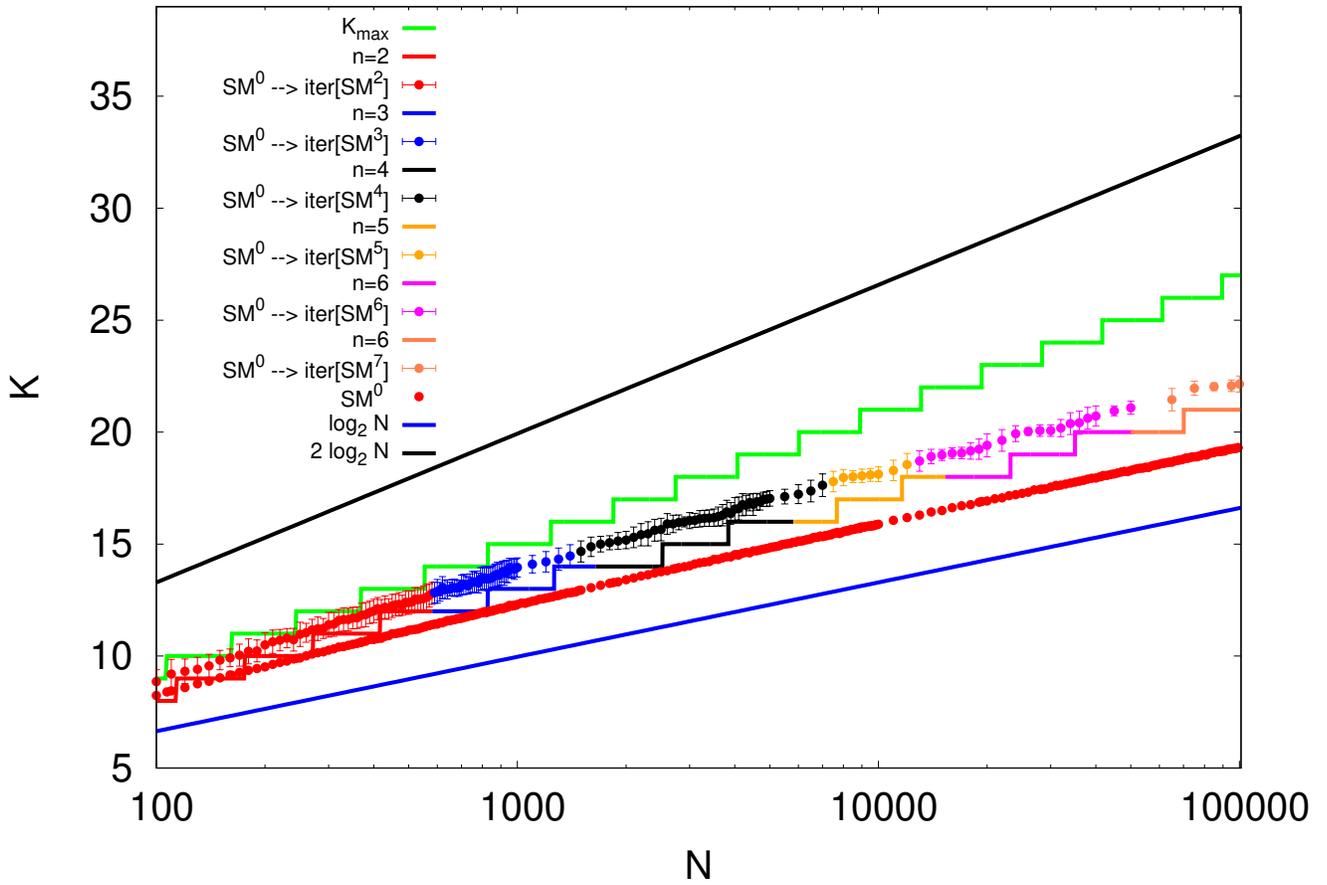}
\caption{Maximal clique sizes found on Erd{\"o}s-R{\'e}nyi graphs $G(N, p=0.5)$.  The green staircase is $K_{\text{max}}$, as in the previous figures. Data points are the                 mean maximal clique size obtained by $SM^{0}\to \text{iter}[SM^{i}] $, using values of $i$ given by Table \ref{Tabi}. The multi-coloured staircase represents the expected maximum values of completed subgraphs conditioned by the fact that we are starting with an arbitrary  complete subgraph of size $i$.}
\label{figSMiterwhichi1}
\end{figure}

We present in Fig. \ref{figSMiterwhichi1} the results of $SM^{0}\to \text{iter}[SM^{i}]$, over the full range of $N$ from $100$ to $100000$.
We use different $i$ in different ranges of $N$, determining their values by experiments. As $N$ increases we have to increase the number of sites kept for the iteration in order to get a bigger complete subgraph at the end of the process. The values of $i$ selected are given in Table \ref{Tabi} 

\begin{table}
\centering
\begin{tabular}{||c | c ||} 
 \hline
$N$ & $i$ \\ [0.5ex] 
 \hline
 100-589 & 2 \\ 
 \hline
 590-1499 & 3\\
 \hline
1500-7499 & 4\\
 \hline
7500-12999 & 5\\
 \hline
13000-64999 & 6\\
 \hline
 65000-100000 & 7\\ [1ex] 
 \hline
\end{tabular}
\caption{The table shows the values of $N$, i.e. the order of the graph, and the values of $i$, the order of the complete sub-graph used as starting point, for the algorithm $SM^{0}\to \text{iter}[SM^{i}]$.}
\label{Tabi} 
\end{table}


Fig. \ref{figSMiterwhichi1} shows the results of experiments with $SM^{0}\to \text{iter}[SM^{i}] $, with $i$ fixed in the range given by Tab. \ref{Tabi}. They fall between two \textit{staircases}.  The upper one is $K_{\text{max}}$, as before, and the lower one is the max clique size predicted by the first moment bound if we begin with a randomly selected clique of size $i$.  The coloured staircase curve is given by \cite{feller1} the smallest value of $K$ for which : 
\begin{equation}
\label{eqcsc}
{N - i \choose K -i }p^{{K\choose 2}-{i\choose 2}} \geq 1.
\end{equation}
This implies that the subgraphs we have selected as a basis for our iteration are much better than average, compared with the very large number of starting subgraphs that a full $SM^i$ would have needed to search. Also, while the slope of the $SM^0$ results is no greater than 1 on this plot at larger values of $N$, the iterated results are rising with a higher slope all the way to $N = 10^5$.

\begin{figure}
\centering
\includegraphics[width=1\columnwidth, keepaspectratio=true, angle=0]{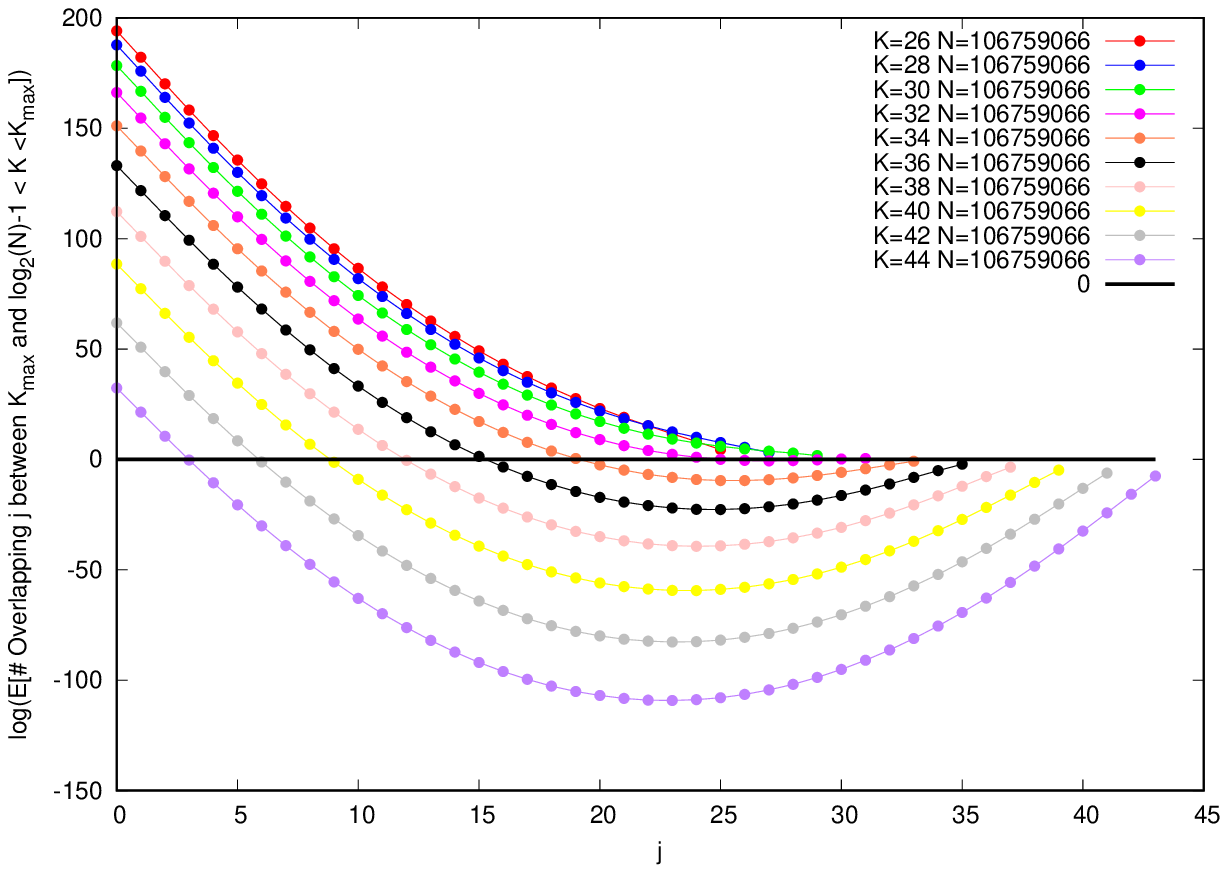}
\caption{Expected number of overlaps between a single clique of size $k < K_{\text{max}}$ and all other cliques of the same size versus $j$, the amount of sites common to the two cliques. For this extreme value of  $N = 106759068 $,  we are at the start of the step in which $K_{\text{max}}$  becomes 46, while $\log_2N$ is 26.7.  A gap has opened at $K = 32$.  For larger values of $K$, overlaps rapidly shrink.  Cliques with $K = 44$ will overlap at three or fewer sites. 
}
\label{overlap}
\end{figure}

To get a more objective standard of difficulty, we use the ideas of the Overlap Gap Property (OGP)
\cite{gamarnik2021overlap}, 
but apply it at finite scales. The expected number of combinations of overlapping clusters in this model is easily calculated. In Fig. \ref{overlap} we show the expected number of cliques of size $K$ that overlap a single randomly chosen clique of the same size on precisely $j$ sites. Values of $K$ range from $\log_2N$ to $K_{\text{max}} - 1$.   We define $K_1$ as the largest value of $K$ for which cliques of size $K$ have some overlaps at all values of $j$. A local search, which moves from one clique to another by changing only one site at a time can still visit all possible cliques of sizes $K_1$ or less. At fixed $N$, cliques larger than $K_1$ will have an overlap gap and will occur only in tiny clusters that differ in a few sites, touching other cliques of the same size only at their edges.  This gap opens up when the overlap $j$ is roughly $\log_2 N$, so we see that $K_1 (N)$ is somewhat larger than $\log_2 N$. 

From plots like Fig. \ref{overlap}, evaluated at the step edges seen for values of $N$ from $10^3$ to $5\,10^4$, we find values of $K_1/ \log_2 N$ decreasing slowly from 1.34 to 1.24. This sets a bound on the values that $SM^{0}$ could discover across this range of $N$. In fact, results of $SM^{0}$ decreased from  $1.2 \log_2 N$ to  $1.17 \log_2 N$ across this range.  However, our results with algorithms $SM^1$ and $SM^2$, seen in Fig. \ref{figSMiterwhichi1}, which reduce confusion by a limited amount of backtracking, exceed it.  At step edges, they range from $1.39$ to $1.36 \log_2 N$.  The prefactor is still decreasing with increasing $N$, but exceeds the suggested OGP limit.

\section{Hidden Clique}

To "hide" a clique for computer experiments, it is conventional to use the first $K_{HC}$ sites as the hidden subset, which makes it easy to observe the success or failure of oblivious algorithms. But this is an entirely different problem than Maximum Clique. Since the hidden clique is unique and distinguished from the many accidental cliques by its greater size, there is no confusion to obstruct the search.

We construct the hidden clique in one of two ways.  The first is simply to restore all the missing links among the first $K_{HC}$ sites.  This has the drawback that those sites will have more neighbors than average, and might be discovered by exploiting this fact. In fact, the upper limit to interesting hidden clique sizes was pointed out by Ku{\v{c}}era \cite{kuvcera1995expected}, who showed that a clique of size $ \alpha \sqrt{N \ln N}$ for a sufficiently large $\alpha$ will consist of the sites with the largest number of neighbors, and thus can be found by $SM^0$.

The second method is to move links around within the random graph in such a way that after the hidden clique is constructed, each site will have the same number of links that it had before.  To do this, before we add a link between sites $i$ and $j$ in the hidden clique, we select at random two sites, $k$ and $l$, which lie outside the clique. $ k$ must be a neighbor of $i$ and $l$ must be a neighbor of $j$.  If $k$ and $l$ are distinct and not neighbors, we create a new link between them, and remove the links between $i$ and $k$ and between $j$ and $l$.  If this fails we try the replacement again, still selecting sites $k$ and $l$ at random.  The result is a new graph with the same distribution of connectivities, as measured from the individual sites.  This sort of \textit{smoothing} of the planting of a hidden clique had been explored by  \cite{sanchis1994test}.  Several graphs prepared in this way are in the DIMACS portfolio, and are said to be more difficult to solve.  In the results we report below, we have used only the first method, as we found no difference resulted when studying hidden cliques in the region of greatest interest, close to the naturally occurring sizes.

\begin{figure}
\centering
\includegraphics[width=1\columnwidth, keepaspectratio=true, angle=0]{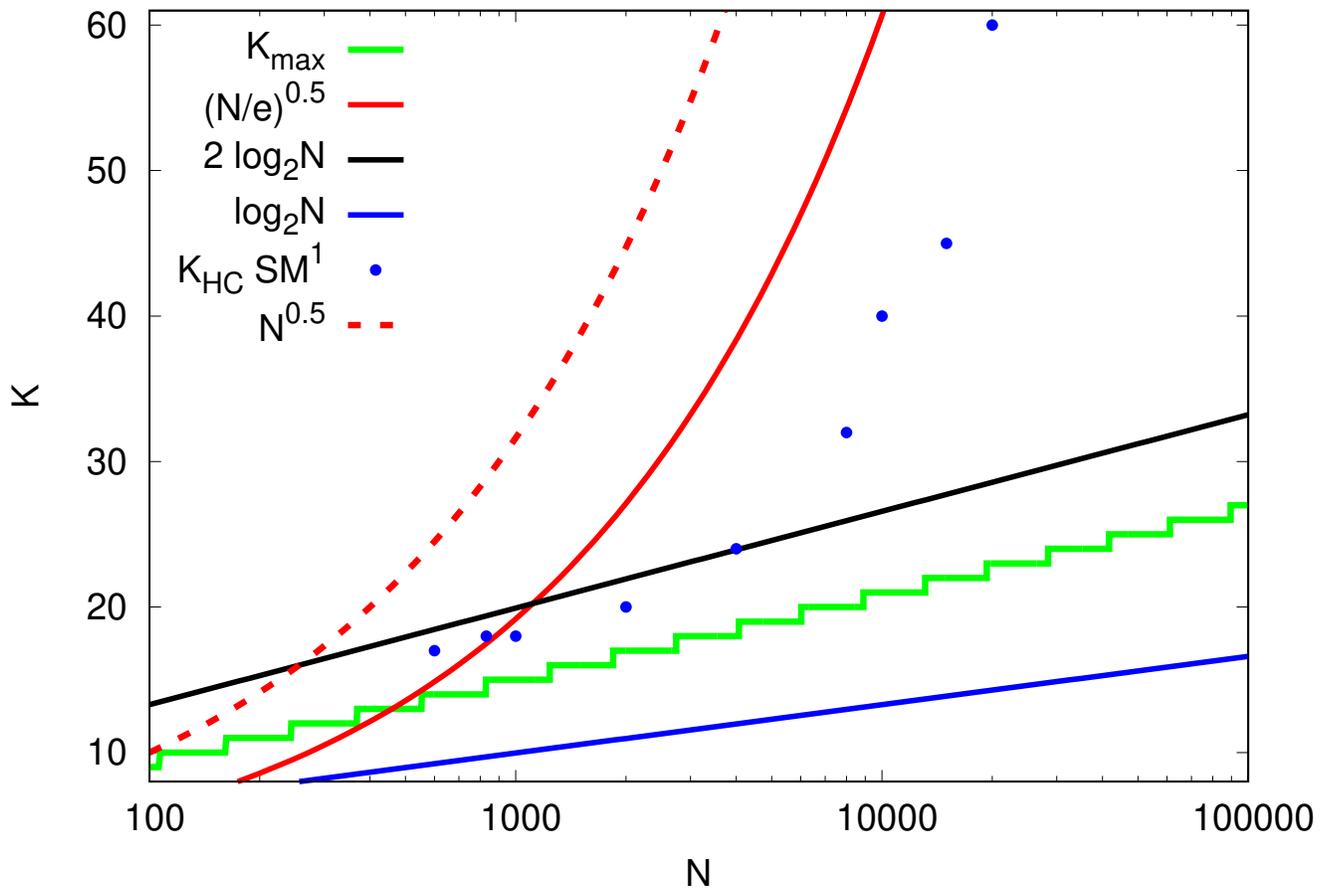}
\caption{ Hidden cliques sizes $K_{HC}$ of interest in $G(N,1/2)$ lie between the $K_{\text{max}}$ staircase and the proven or experimentally observed lower limits that can be found with spectral methods (dashed red line) or AMP techniques  (solid red line).  Sizes of the smallest hidden cliques identified with the $SM^{1}$ method, augmented by early stopping, are shown with blue dots and lie well below both limits.}
\label{fig-mont}
\end{figure}

A stronger result, by Alon et al. \cite{alon1998finding} uses spectral methods to show that a hidden clique, $C$, of cardinality  $|C|\ge 10 \sqrt{N}$ can be found with high probability, in polynomial time. Dekel et al. \cite{dekel2014finding} showed that with a linear ($\mathcal{O}( N^2)$, the number of links) algorithm the constant can be reduced to $1.261$. Our experiments, using this approach, were successful to a slightly lower value, of roughly $1.0$.

Finally, recent work of Deshpande and Montanari \cite{deshpande2015finding} has shown that Approximate Message Passing (\textbf{AMP}), a form of belief propagation, can also identify sites in the hidden clique.  This converges down to $\sqrt{N/e}$, where $e$ is Euler's constant (see Fig. \ref{fig-mont}).  No algorithm currently offers to find a clique of size less than $\sqrt{N/e}$  and bigger than $K_{\text{max}}$, in quasi-linear time, for arbitrary $N$.   Parallel Tempering enhanced with an early stopping strategy \cite{chiara2018parallel, PhysRevE.100.013302}, is able to explore solutions below $\sqrt{N/e}$, but only at the expense of greatly increased computational cost.  Some of these procedures identify some, but perhaps not all of the planted clique sites, and require some ``\textit{cleanup}'' steps to complete the identification of the whole clique. The cleanup procedures all require starting with either a subset of the hidden clique sites and finding sites elsewhere in the graph that link to all of them, or eliminating the sites in a possible mixed subset of valid and incorrect choices which do not extend as well, or doing both in some alternating process.  These can be proven to work if the starting point is sufficiently complete (hence Alon et al.'s $C = 10$ starting point).  We find experimentally, and discuss below, that a cleanup process can be effective given a much poorer starting point.

\subsection{Iterative methods}

Next, we consider methods of searching for the hidden clique that involve iteration. We shall employ two approaches, the $SM^1$ greedy algorithm with a simple modification, and the belief propagation scheme introduced by Deshpande and Montanari \cite{deshpande2015finding}.  First, we must make a further modification of the adjacency matrix.  We will use $\mathbf{\tilde{A}}$, whose elements are defined by: 

\begin{equation*}
\begin{split}
&\tilde{A}_{ij}  =  { \tilde{a}_{ij}  },\\
& \tilde{a}_{ij} = \tilde{a}_{ji} ,
\end{split}
\end{equation*}

where $\tilde{a}_{ij}=1$  if the link is present,  $\tilde{a}_{ij}=-1$  if the link is absent, and   $\tilde{a}_{ii}=0$.

The reason for the extra nonzero entries is simple.  It generates the same band of eigenvalues, with doubled width, and moves the special uniform state at $0.5 \sqrt{N}$ into the center of the band, where it no longer interferes with the influence of eigenstates at the top of the energy band, which are most likely to contain the hidden clique sites. 


Now we considered the  \textbf{AMP} algorithm introduced by Deshpande et al. in \cite{deshpande2015finding}. They proved that as $N \to \infty$ their algorithm is able to find hidden cliques of size $K_{HC} \ge \sqrt{N/e}$ with high probability. 
 \textbf{AMP} is derived as a form of \textit{belief propagation} (\textbf{BP}), a heuristic machine learning method for approximating posterior probabilities  in graphical models. \textbf{BP} is an algorithm \cite{montanari2007solving,felzenszwalb2006efficient,yedidia2001generalized}, which extracts marginal probabilities for each variable node on a factor graph. It is exact on trees, but was found to be effective on loopy graphs as well \cite{deshpande2015finding,mezard2009information,frey1998revolution,mooij2005sufficient}. It is an iterative message passing algorithm that exchanges messages from the links to the nodes, and from them it computes marginal probabilities for each variable node.  When the marginal probability has been found, as \textbf{BP} has converged, one can obtain a solution of the problem, sorting the nodes by their predicted marginal probabilities. However, it is possible, if the graph is not locally a tree, that \textbf{BP}  does not find a solution or converges to a random and uninformative fixed point. In these cases, the algorithm fails.  \textbf{BP} for graphical models runs on factor graphs where each variable node is a site of the  original graph $G(N,p)$, while each function node is on a link of the original graph $G(N,p)$.
 Here we describe briefly the main steps that we have followed in implementing \textbf{AMP} algorithm.  For details, we refer the reader to \cite{deshpande2015finding}.
 \textbf{AMP} runs on a complete graph described by an adjacency matrix  $\mathbf{\tilde{A}}$. \textbf{AMP}  iteratively exchanges messages from links to nodes, and from them, it computes quantities for each node. These quantities represent the property that a variable node is, or not, in the planted set. It is intermediate in complexity and compute cost between local algorithms, such as our greedy search schemes, and global algorithms such as the spectral methods of Alon et al. \cite{alon1998finding}
For our purpose, we implemented a simple version of this algorithm, using Deshpande et al's \cite{deshpande2015finding} equations. Here, we recall them:

\begin{equation}
  \label{messages}
    \Gamma^{t+1}_{i \to j}=\log \frac{K_{HC}}{\sqrt{N}}+\sum^{N}_{l\neq i,j}\log\left( 1+\frac{(1+\tilde{A}_{l,i}) \text{e}^{\Gamma^{t}_{l\to i}}}{\sqrt{N}} \right) -\log\left( 1+\frac{\text{e}^{\Gamma^{t}_{l\to i}}}{\sqrt{N}}\right),
  \end{equation}

\begin{equation}
  \label{marginal}
    \Gamma^{t+1}_{i}=\log \frac{K_{HC}}{\sqrt{N}}+\sum^{N}_{l\neq i}\log\left( 1+\frac{(1+\tilde{A}_{l,i}) \text{e}^{\Gamma^{t}_{l\to i}}}{\sqrt{N}} \right)-\log\left( 1+\frac{\text{e}^{\Gamma^{t}_{l\to i}}}{\sqrt{N}}\right).
  \end{equation}

Equations (\ref{messages}) and (\ref{marginal}) describe the evolution of messages and vertex quantities $\Gamma^{t}_{i}$. They run on a fully connected graph, since both the presence or absence of a link between sites is described in the adjacency matrix $\mathbf{\tilde{A}}$. For numerical stability, they are written using logarithms. Initial conditions for messages in  (\ref{messages}) are randomly distributed and less than $0$.
 The constant part is obtained by observing that relevant scaling for hidden clique problems is $\sqrt{N}$.  
 
 Equation (\ref{messages}) describes the numerical updating of the outgoing message from site $i$ to site $j$. It is computed from all ingoing messages to $i$, obtained at the previous iteration, excluding the  outgoing message from $j$ to $i$. These messages, i.e. equation (\ref{messages}),  are all in $ \mathbb{R}$ and they correspond to so-called odds ratios that vertex $i$ will be in the hidden set $C$. In other words, the message from  $i$ to $j$ informs  site $j$ if site $i$ belongs to the hidden set or not, computing the odds ratios of all remaining  $N-2$ $l$ sites of the graph, with $l\neq i,j$. When a site $l$ is connected to site $i$, the difference between logarithms, in the sum,  will be positive and will correspond to the event that the site $l$ is more likely to be a site of $C$ than a site outside it.  However, when $l$ is not connected to $i$ the corresponding odds ratios will be less than one, i.e. the difference of logarithms, in the sum of equation (\ref{messages}),  will be less than zero, and will correspond to the event that the site $l$ is more likely to be outside the hidden set. The sum of all the odds ratios will update equation (\ref{messages}),  telling us  if site $i$ is more likely to be in $C$ or not.
 
 Equation (\ref{marginal}), instead, describes the numerical updating of the vertex quantity  $\Gamma^{t}_{i}$. It is computed from all ingoing messages in $i$, and  is an estimation of the likelihood that $i\in C$. These quantities are larger for vertices that are more likely to belong to the hidden clique \cite{deshpande2015finding}. Elements of the hidden set, therefore, will have $\Gamma^{t_c}_{i\in C}>0$, while  elements that are not in the hidden set will  have  $\Gamma^{t_c}_{i\not \in C}<0$.

 As iterative \textbf{BP}  equations,  (\ref{messages}) and (\ref{marginal})  are useful only if they converge. The computational complexity of each iteration is $\mathcal{O}(N^2)$, indeed, equation (\ref{messages}) can be computed efficiently using the following observation:
 
 \begin{equation}
   \label{trick}
     \Gamma^{t+1}_{i \to j}=\Gamma^{t+1}_{i}-\log\left( 1+\frac{(1+\tilde{A}_{j,i}) \text{e}^{\Gamma^{t}_{j\to i}}}{\sqrt{N}} \right)+\log\left( 1+\frac{\text{e}^{\Gamma^{t}_{j\to i}}}{\sqrt{N}}\right).
   \end{equation}
 
 The number of iterations needed for convergence for all messages/vertex quantities is of order $\mathcal{O}(\log N)$, which means that the total computational complexity of the algorithm is $\mathcal{O}(N^2 \log N)$. Once all messages in (\ref{messages}) converge, the vertex quantities given by (\ref{marginal}) are sorted into descending order. Then, the first $K_{HC}$ components are chosen and checked to see if they are a solution. If a solution is found we stop with a successful assignment, else the algorithm returns a failure. For completeness, our version of the\textbf{AMP} algorithm  returns a failure also when it does not converge after $t_{\text{max}}=100$ iterations.

 As a first experiment, we run simulations, which reproduce the analysis in \cite{deshpande2015finding}, but apply their methods to a larger sample, $N = 10^4$.  In Fig. \ref{AMPVSSM1} we show the fraction of successful recoveries by \textbf{AMP} after one convergence, as a function of $\alpha$. As the analysis in \cite{deshpande2015finding} predicts, the  \textbf{AMP} messages converge down to about $\alpha = \sqrt{N/e}$, but with a decreasing probability of convergence, or with success in a decreasing fraction of the graphs that we have created.  At and below the algorithmic threshold of \textbf{AMP} for this problem, we obtained very few solutions.

 \subsection{Greedy search with early stopping}


\begin{figure}
\centering
\includegraphics[width=1\columnwidth, keepaspectratio=true, angle=0]{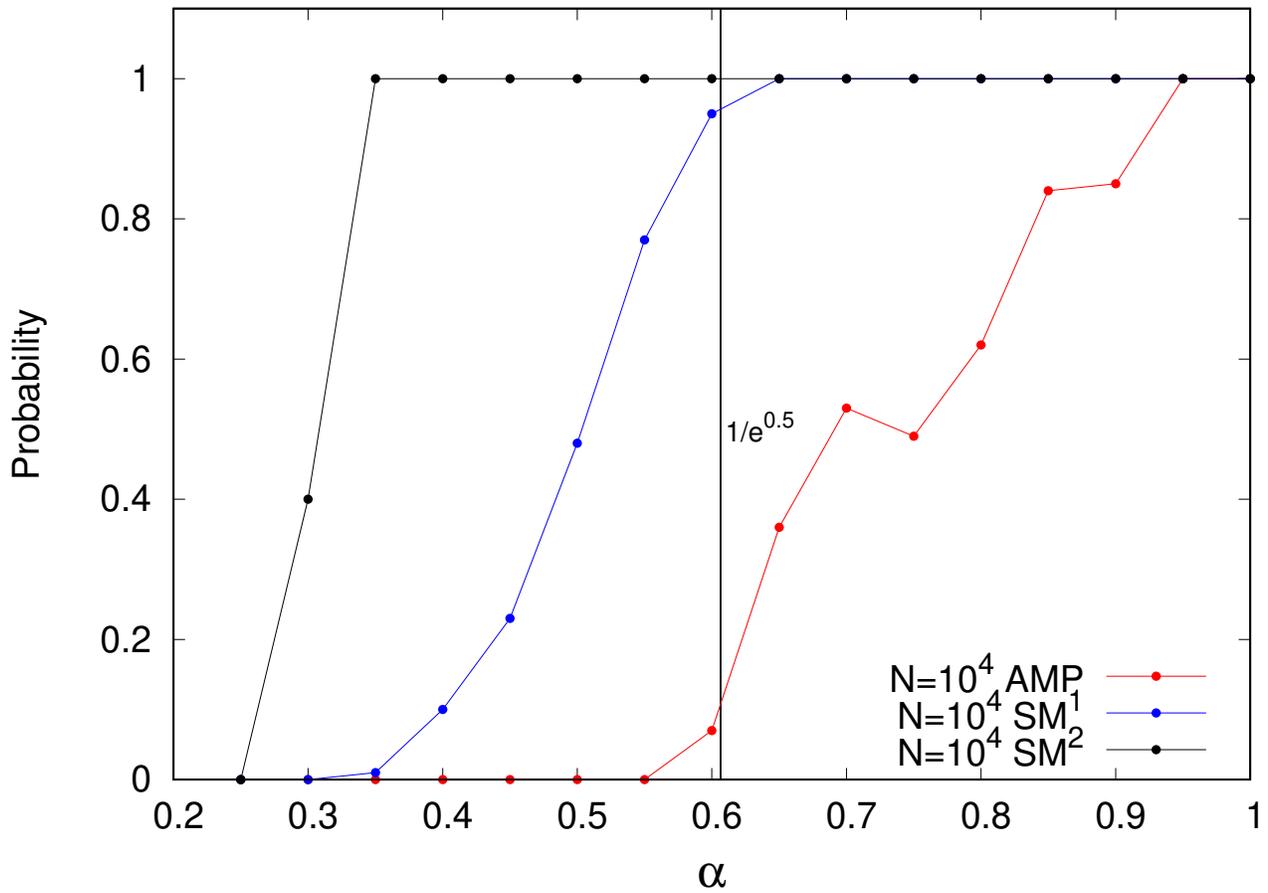}
\caption{The picture shows the probability of success for recovering the hidden clique as a function of $\alpha$ for \textbf{AMP} (red points), $SM^1$ with early stopping (blue points) and $SM^2$ with early stopping (black points). 
Points are averages over $100$ graphs $G(N=10^4, p=0.5, K_{HC}=\alpha \sqrt{N})$ for \textbf{AMP} algorithm and $SM^1$ with early stopping, while they are an average over $5$ graphs for $SM^2$ with early stopping. }
\label{AMPVSSM1}
\end{figure}

\begin{figure}
\centering
\includegraphics[width=1\columnwidth, keepaspectratio=true, angle=0]{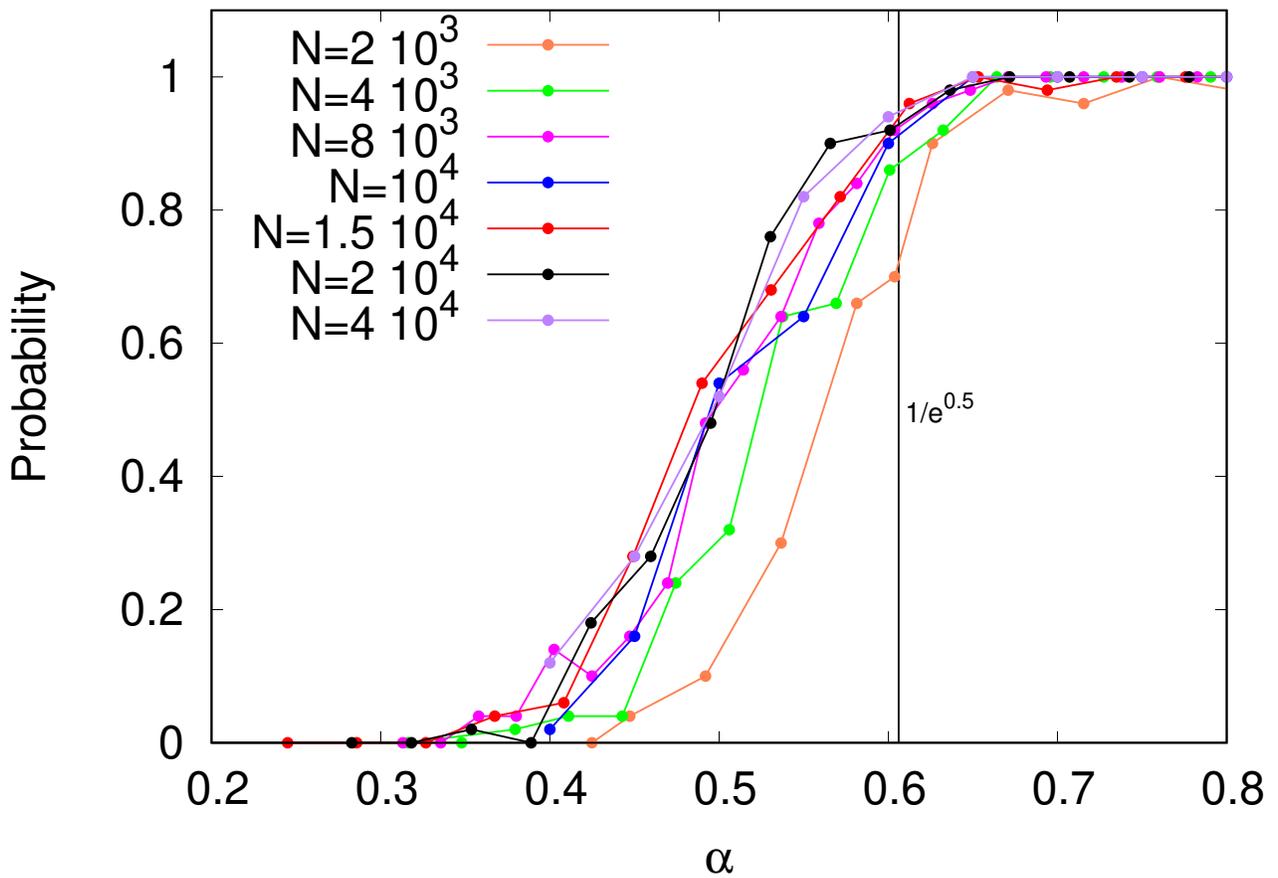}
\caption{Probability of solution vs $\alpha$  using SM[1] with early stopping, at graph sizes from 2000 to 40000. }
\label{fig-Pvsalpha}
\end{figure}


\begin{figure}
\centering
\includegraphics[width=1\columnwidth, keepaspectratio=true, angle=0]{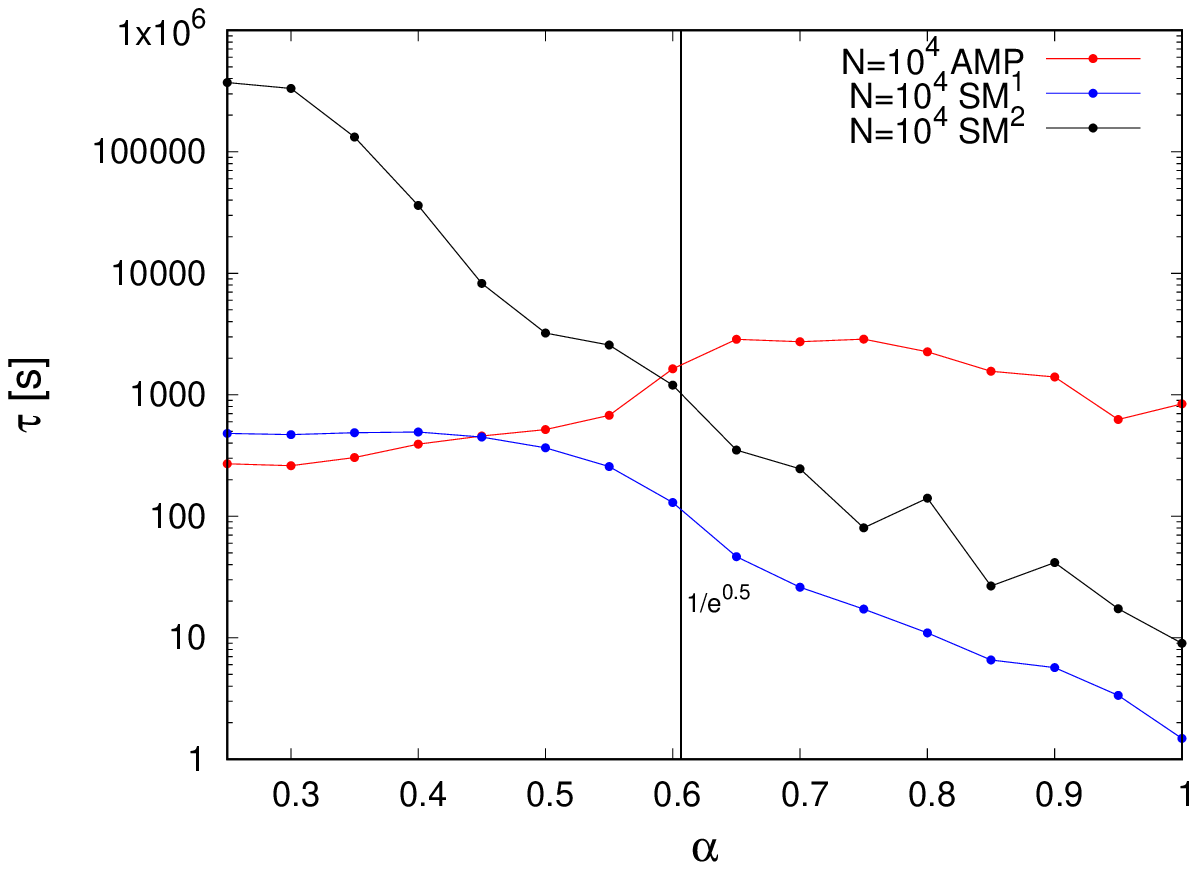}
\caption{The picture shows as a function of $\alpha$ the average time, in seconds, needed by \textbf{AMP} (red points), and $SM^1$ (blue points) and  $SM^2$ (black points) with early stopping, to give an answer on a single graph $G(N=10^4, p=0.5, K_{HC}=\alpha \sqrt{N})$.}
\label{TIME}  
\end{figure}

We also explored using our greedy search methods to uncover a planted clique in this difficult regime.  Our hypothesis was that using $SM^0$ was unlikely to succeed since almost all sites selected at random do not lie within the planted clique.  But $SM^1$ seems more promising, even with its $\mathcal{O}(N^3)$ cost.  And if the search gave rise to any clique of size $R(N)$ or larger, perhaps by a fixed amount $d \ge 2$,  that is strong evidence of the existence of the planted clique. A clique of this size is a reliable starting point for a cleanup operation to find the remaining sites.

The cleaning algorithm starts with a complete subgraph $C$ of order $|C|=R(N)+2$, obviously too large to be just any statistically generated clique.   We scan the entire remaining graph, selecting the sites with the most links to the subset  $C$, and adding them to $C$ to form a subset $C'$.  The largest clique to be found in $C'$ will add new sites not found before and may lose a few sites which did not belong.  Iterating this process a few times until no other sites can be added, in practice, gives us the hidden clique.

 As shown in Fig. \ref{AMPVSSM1}, this succeeds in a greater fraction of the graphs than does \textbf{AMP} for planted cliques, when $\alpha < 1$.  This strategy of stopping $SM^1$ as soon as the hidden clique is sufficiently exposed to finish the job
produces the hidden clique almost without exception in our graphs of order $10^4$ through the entire regime from $\alpha = 1$  down to $\alpha = 1/\sqrt{e}$.  In this regime, \textbf{AMP},  converges to a solution in a rapidly decreasing fraction of the graphs.  We studied the same $100$ graphs with $\mathbf{AMP}$ as were solved with $SM^1$ at each value of $\alpha$.  Using $SM^1$ with early stopping, we could extract planted cliques as small as $\alpha = 0.4$.

In Fig. \ref{fig-Pvsalpha}, we show the probability of success for a range of values of N.  Except for the two cases for $N = 200$ or $400$, where the solutions lie close to the naturally generated clique sizes, the success probability curves track the results obtained at $N = 10^4$.   Notice how the curves show less scatter as $\alpha$ increases.

The success of early stopping in making $SM^1$ useful led us to try the same with $SM^2$.  We tried this with only $5$ graphs at each value of $\alpha$, and were able to identify the planted clique in all graphs down to $\alpha = 0.35$, and in two out of five graphs at $\alpha = 0.3$.  The method was not successful at all at $\alpha = 0.25$.

 Fig. \ref{AMPVSSM1} compares the results of all three methods on our test case $N = 10^4$.  It appears that the \textit{local}, greedy methods, when used repeatedly in this fashion, are actually stronger than the more globally extended survey data collected by \textbf{AMP}.  But to compare their effectiveness, it is also necessary to compare their computational costs. This is explored in Fig.  \ref{TIME}.


In Fig. \ref{TIME}, we compare the effectiveness of $SM^1$ and $SM^2$ with early stopping and \textbf{AMP}.  First, we find that the number of trials required for $SM^1$ to expose the hidden clique was close to $N$ at the lowest successful searches, but dropped rapidly (the scale is logarithmic) for $\alpha > \sqrt{1/e}$.  As $\alpha$ approaches $1$, there are more starting points than there are points in the hidden clique, while for $\alpha < \sqrt{1/e}$, not every point in the hidden clique is an effective starting point. 

We briefly explored the importance of where to stop the search by running $SM^1$ to completion for a small number of graphs at $\alpha = 0.6$ and considering the sizes of the cliques found.  This distribution varies quite widely from one graph to another.  The full hidden clique is frequently found, and the most common size results were about half of the hidden clique size.  Only a very few cliques returned by $SM^1$ were within $1-4$ sites of $R(N)$, so we recommend the stopping criterion $R(N) + 2$ as a robust value.

The average running time to solve one graph for each of the three is plotted in Fig. \ref{TIME}. The average cost of solving \textbf{AMP}, (red points) is greatest just above $\alpha = 1/\sqrt{e}$ where it sometimes fails to converge, and decreases at higher $\alpha$, largely because convergence is achieved, with fewer iterations as $\alpha$ increases.  The cost decreases at lower values of $\alpha$ because \textbf{AMP} converges more quickly, but this time to an uninformative fixed point.   $SM^1$ with early stopping (blue points) requires less time than \textbf{AMP} to expose the planted clique at all values of $\alpha$ where one or both of the methods are able to succeed and is several hundred times faster at $\alpha = 1$.   $SM^2$ with early stopping (black points) is more expensive than  $SM^1$ with early stopping at all values of $\alpha$, but is also less costly than \textbf{AMP} in the range $1/\sqrt{e} < \alpha <1$.  It is the most costly algorithm at still lower values of $\alpha$, but the only method that can provide any solutions down to the present lower limit of $\alpha = 0.3$.  This efficiency, as well as the ability of local greedy algorithms with early stopping to identify cliques with $\alpha < \sqrt{1/e}$, is a surprising and novel result.





\section{Conclusions}
\label{sec:Conclusions}

More than $30$ years have elapsed since the DIMACS community reviewed algorithms for finding maximum cliques (and independent sets) in Erd{\"o}s-R{\'e}nyi graphs $G(N,p)$ with $N$ sites and bonds present with fixed probability, $p$.  Computer power and computer memory roughly $100\times$ what was available to the researchers of that period are now found in common laptops. But unfortunately, the size of the problems that this can solve (in this area) only increases as the $\log$ or a fractional power of the CPU speed.
We can now explore the limits of polynomial algorithms up to $N=10^5$, while the DIMACS studies reached only a few thousand sites.  In contrast to problems like random $3$-SAT, for which almost all instances have solutions by directed search \cite{selman1993local} or belief propagation-like \cite{mezard2002analytic,mezard2007constraint,marino2016backtracking} methods which approach the limits of satisfiability to within a percent or less, finding a maximum clique remains hard over a large region of parameters for almost all random graphs, if we seek solutions more than a few steps beyond $\log_2 N$.  Using tests more detailed than the \textit{bakeoff}  with which algorithms have been compared, we show that expensive $\mathcal{O}(N^3)$ and $\mathcal{O}(N^4)$ searches can accurately reproduce the distribution of maximum clique sizes known to exist in fairly large random graphs.  (Up to at least $N = 500$ for the $\mathcal{O}(N^3)$ algorithm and about $N = 1500$ for the $\mathcal{O}(N^4)$ algorithm.) This is a more demanding and informative test of the algorithms' performance than seeing what size clique they each can extract from graphs whose actual maximum clique size is unknown.

A more promising approach is to use the simplest search algorithm to define a subgraph much smaller than $N$ as a starting subset in which to apply the higher-order search strategies. This does not produce the exact maximum clique, or even get within a percent or less of the answer as with SAT, because the naive initial search combines sites which belong in different maximum cliques into the starting set. The higher-order follow-up search that we employ does not fully separate them. Therefore, from our initially defined clusters, which exceed the $\log_2 N$ lower bound in size, but may contain a confusing mixture of incompatible larger clusters that prevent each other from being extended we have used a restricted form of backtracking to find the subset which is most successful in growing further.  

The second challenge we considered is locating and reconstructing a hidden clique.  Using Deshpande and Montanari's \cite{deshpande2015finding} \textbf{AMP}, or our slightly more than linear cost polynomial $SM^1$ with early stopping, we can discover and reconstruct the hidden clique well below the limit of spectral methods and almost down to the sizes of naturally occurring cliques.  It is surprising that a version of the simplest greedy algorithm performs better on problems of the largest currently achievable sizes. This is possible because the hidden clique is unique, so the greedy search is not confused.

The challenges posed at the start of this paper apply only in the limit  $N \to \infty$, in a problem with significant and interesting finite-size corrections.  Although computing power, data storage, and the data from which information retrieval tools are sought to find tightly connected communities all increase at a dramatic pace, all of these presently lie in the  \textit{finite-size} range of interest, far from an asymptotic limit. Yet they are well beyond the scale of previous efforts to assess algorithms for this problem.  Since asymptotic behavior is  only approached logarithmically in the clique problem, we think that further progress is possible in the finite-size regime.  We have shown that effective searches for cliques can be conducted on graphs of up to $10^5$ sites, using serial programs.  With better, perhaps parallel algorithms, and the use of less-local search strategies such as \textbf{AMP}, can this sort of search deal with information structures of up to $10^9$ nodes using today's computers?  With computational resources of the next decade, and perhaps a better understanding of the nature of search in problems with such low signal-to-noise ratios as Maximum Clique, perhaps we can hope to see graphs of order the earth's population being handled.

What we learn from looking into the details of the Overlap Gap
\cite{gamarnik2021overlap}
(Fig. \ref {overlap} and many similar plots) is that an ergodic phase in which local rearrangements can transform
any clique of size $K_1$ into another of the same size extends slightly above $K = \log_2 N$. But cliques of size $K_1$ vastly outnumber cliques of size $K_{\text{max}}$ and the larger cliques only slightly overlap, so what is needed are efficient ways of identifying the components of such cliques with the greatest chance of continuing to grow by local moves. 

Beyond the Overlap Gap, we expect to encounter a "Clustered phase" similar to that predicted for more complex systems using the arguments of replica symmetry breaking and the calculational tools of the "cavity model"\cite{krzakala2007gibbs, zdeborova2008statistical}. (Fortunately, the extra complications of "frozen variables," introduced in these models, should be absent when looking at graph properties such as Maximum Clique, where there are no extra internal degrees of freedom.)

We have made progress beyond the Overlap Gap limit by simple forms of backtracking. More expensive searches, costing higher powers of $N$, such as $SM^1$ and $SM^2$ are effective for a range of $N$ within this clustered phase. But stronger and more costly methods, such as simulated annealing\cite{kirkpatrick1983optimization} and its parallel extensions, or learning algorithms \cite{bengio2021machine} may be needed to provide more exhaustive search.

\section*{Data availability statement}
The numerical codes used in this study and the data that support the findings are available from the corresponding author upon request.

\bibliography{sample}

\section*{Acknowledgements}
RM and SK were supported by the Federman Cyber Security Center of the Hebrew University of Jerusalem.  Continued collaboration was hosted at the University of Rome "La Sapienza", and at the MIT Media Lab.  We enjoyed stimulating conversations with Federico Ricci-Tersenghi and Maria Chiara Angelini. RM thanks  the FARE grant No. R167TEEE7B and the grant "A multiscale integrated approach to the study of the nervous system in health and disease (MNESYS)".

\section*{Author contributions statement}
All authors contributed to all aspects of this work.

\end{document}